\documentclass[twocolumn]{aastex701}

\usepackage{upgreek}
\usepackage{chemformula}
\usepackage[caption=false]{subfig}
\usepackage{mathtools}

\usepackage{xcolor}
\usepackage{soul}
\definecolor{turquoisehl}{rgb}{0,1,1}
\definecolor{greyhl}{rgb}{0.8,0.8,0.8}
\definecolor{redhl}{rgb}{1,0.5,0.5}
\definecolor{bluehl}{rgb}{0.75,0.75,1}
\definecolor{greenhl}{rgb}{0.5,1.0,0.5}
\definecolor{purplehl}{rgb}{0.75,0.5,0.9}
\definecolor{orangehl}{rgb}{1,0.4,0.3}
\definecolor{pinkhl}{rgb}{0.9,0.4,0.9}

\graphicspath{ {./figures/} }

\submitjournal{The Planetary Science Journal} 

\shorttitle{Mercury's Radiation-hard Sulfides}
\shortauthors{J\"aggi et al.}

\begin{document}

%
%


\title{\added{Surface Response of Mercury’s Sulfides under Solar Wind Ion Irradiation}}

%
%




\author[0000-0002-2740-7965]{Noah Jäggi}
\affiliation{Laboratory for Astrophysics and Surface Physics, University of Virginia, 395 McCormick Road, Charlottesville, VA 22904, USA}
\email[show]{noah.jaeggi@bluewin.ch}

\author[0000-0002-2740-7965]{Catherine A. Dukes}
\affiliation{Laboratory for Astrophysics and Surface Physics, University of Virginia, 395 McCormick Road, Charlottesville, VA 22904, USA}
\email{cdukes@virginia.edu}



\begin{abstract}
The MESSENGER mission revealed unexpectedly high sulfur content within Mercury's surface, deviating from the Lunar regolith, which was—until recently—considered a good Mercury analogue. 
Mercury’s exposure to energetic space weathering processes such as meteoritic impact and solar-wind sputtering suggests this high sulfur concentration should be reflected in the suprathermal sulfur population of the Hermean exosphere. UV spectroscopy has not yet detected exospheric sulfur, a result attributed primarily to its low glow-factor. Future detection by BepiColombo’s Mass Spectrum Analyzer depends on sulfur abundance in the exosphere.  
Radiation-induced segregation has been observed in the common sulfide troilite (FeS), a constituent mineral in returned Lunar samples, meteorites, and asteroids, where the resulting metal cap is expected to reduce sulfur ejection to Mercury's exosphere. 
In this work, we investigate the irradiation response of Mercury-relevant sulfides. Niningerite (MgS) and oldhamite (CaS) were irradiated with solar-wind speed 2~keV~\ch{H_2+} or 4~keV~\ch{He+}, and in-situ compositional and chemical bond analysis as a function of fluence was performed using an XPS microprobe.   
Neither MgS nor CaS expressed detectable damage-induced segregation and instead reached metal-to-sulfur ratios close to bulk with irradiation.   
Based on this finding, structural information, and literature analyses, we infer that an S–S anionic spacing exceeding $\sim$3.2~\AA{} inhibits radiation-induced sulfur depletion and promotes stoichiometric sputtering.  
We therefore predict no cation (metal) surface segregation in Hermean sulfides and no reduction in suprathermal sulfur emission caused by metal cladding formation in TiS, CrS, and Ca–Mg sulfides. This radiation-hardness for Mercury-relevant sulfides is novel and unexpected, and should facilitate detection in Mercury's exosphere by the BepiColombo mission.
  
\end{abstract}

\section{Introduction}

Solar wind ion sputtering and micrometeoroid impact vaporization are the sourcing mechanisms for supplying heavy elements such as Ca, Mg, and S to the suprathermal exospheric ion population of Mercury. Experimental work on FeS has shown that the sputtering yield of S is expected to significantly decrease after \ch{H2+} and \ch{He+} irradiation due to the rapid formation of a Fe-metal capping layer on the sulfide surface\citep{Christoph2022}. The apparent S-depletion in comparison to ordinary chondritic meteorites observed on asteroids such as 433~Eros was explained with this ion-surface interaction, along with micrometeorite impact \citep{nittler_x-ray_2001,killen_depletion_2003,kracher_space_2005,loeffler_laboratory_2008}. This raises the questions: do Mercury's sulfides show the same surface sulfur depletion behavior? Is there a universally reduced S yield for all irradiated sulfides as suggested in \cite{Jaggi2024}?

\subsection{The missing exospheric sulfur}
Based on MESSENGER observations, there is a stark discrepancy between the amount of sulfur detected on the surface of Mercury compared to its exosphere. Mercury's regolith exhibits a relatively high sulfur content \citep[$\leq$3.5 wt\%;][]{McCoy2018,Cartier2019} compared to the Moon \citep[max. of 0.21~wt\% found in Luna 16 soils][]{Heiken1991} and there are strong local abundance correlations of S with Mg in the Intercrater Plains and Heavily Cratered Terrain \cite{Weider2012} and of S with Ca across Mercury’s surface \citep{Nittler2011,Weider2012}, based on data collected by MESSENGER’s X-ray Spectrometer (XRS) and Gamma-Ray and Neutron Spectrometer (GRNS).
In the exosphere, out of the three elements in Mg-Ca sulfides, only Ca and Mg have been successfully detected via UV-VIS optical detection \citep{Doressoundiram2009,McClintock2009} whereas S abundance has so far only been constrained to an upper limit \citep{Grava2021}. The explanation for this is either a low S concentration in the exosphere, the consistently low g-value---the photon emission probability per excited atom, note that the g-value for sulfur is about 100x lower than sodium---of S, or both \citep{Grava2021,Killen2022b}. We therefore wouldn't expect optical detection to be the preferred method of sulfur observation. We would rather expect that S will be detected by the Mass Spectrum Analyzer on board of BepiColombo \citep{Delcourt2016}. The FIPS aboard Messenger did not have sufficient mass resolution necessary to clearly distinguish elements in the range 32-35 amu to definitively separate S+, from O2+ and/or H2S+\citep{Zurbuchen2008,raines_distribution_2013,Delcourt2016}. The Mass Spectrum Analyzer aboard BepiColombo, however, boasts a mass resolution (m/$\Delta$m) of 40, which will suffice to distinguish mass 32 from other atomic ions.

\subsection{Origin of Sulfides on Mercury}
The presence and stability of sulfide minerals is possible due to the reducing environment present on/within Mercury's surface at present and during sulfide formation. Mercury exhibits an unusually low average Fe content \citep[$\sim$1.5~wt\%;][]{Nittler2011,evans_major-element_2012,Weider2015} in addition to the relatively high S content. These elemental abundances allow for the estimation of Mercury’s oxygen fugacity \citep[f\ch{O2}; e.g.,][]{Albarede2011}, which reflects the availability of oxygen for chemical reactions during planetary formation and thus indicates the planet’s redox state. Assuming all surface Fe is oxidized, the calculated f\ch{O2} would be 2.8 to 4.5 log units below the iron-wüstite (IW) buffer at 0.1 MPa, based on thermodynamic models \citep{Cartier2019,Zolotov2013,Robie1995}. Alternatively, using sulfur as an oxybarometer \citep{Namur2016a} yields a mean f\ch{O2} of IW-5.4, suggesting that Mercury is more reduced than the Moon \citep[IW-2 to IW;][]{Karner2006,Wadhwa2008}, and even the most reduced meteorites, such as enstatite chondrites \citep[IW-5;][]{Rubin1988,Wadhwa2008}. This discrepancy between f\ch{O2} values derived from Fe and from S was suggested to be due to Fe being hosted in sulfide phases rather than within silicates.

The highly-reduced nature of Mercury is the basis to explain how large quantities of sulfides can end up in Mercury's regolith. On one hand, a reducing environment would allow CaS to remain stable in a melt before crystallizing in effusive melt layers during the last planetary volcanic period \citep{anzures_oxygen_2025}, or there might have been a resurfacing of an early sulfide crust that was formed under reducing conditions \citep{boukare_production_2019,marchi_global_2013,Weider2012} with a CaS reservoir inherited from enstatite chondrite building blocks \citep{hammouda_message_2022,lodders_experimental_1996,floss_rare_1990}. On the other hand, \cite{Renggli2023} has shown that MgS, CaS, and more complex sulfides form by the reaction of a reduced S gas with glass or rock forming minerals (olivine, pyroxene, and plagioclase). This sulfidation process might be responsible for the presence of surface sulfides, even after the last period of extrusive volcanism by gas escaping along faults from volatile-rich lower level into volatile-free upper regolith \citep{Rodriguez2020}. 

\subsection{Radiation Induced Segregation}
The surface composition of a sample altered by ion irradiation is defined by the combined effect of preferential sputtering, implantation, and transport processes \citep{Behrisch1983}. In the case of most rock-forming silicates, the surface will preferentially lose the lightest constituents with the lowest surface binding energy. This depletion causes a relative accumulation of the other species which will subsequently cause an increase in yield of the now more abundant constituent until the yield composition eventually reaches an equilibrium and becomes approximately stoichiometric to the pristine bulk. With H and He irradiation, atomic displacements in the material lattice are created by impactor collision cascades, which saturate in the top tens of nm in rock forming minerals \citep[e.g.,][]{Jaggi2021a}. 

The process of element segregation induced by ion impacts has been known for over 50 years \citep{kelly_bombardment-induced_1989,kelly_summary_1985} and was re-branded from ‘bombardment-induced segregation' in the late 80s to ‘radiation-induced segregation' (RIS) in the 2010s \citep{Naguib1975,nastar_118_2012,wang_radiation-induced_2020}. The process of species segregation is well understood in metal alloys \citep{nastar_118_2012} and is related to the production rate and mobility (or diffusivity) of defects compared to the rate of elimination of excess point defects. More recently, this link was also made for oxides \citep{wang_radiation-induced_2020}. To the best of our knowledge, there has not been a systematic, conclusive description for how the chemical and composition of rock-forming sulfide minerals (e.g., CaS, MgS, TiS, ZnS, and CuFeS2) reacts to solar wind irradiation. 

Mobilization of S in troilite (\ch{FeS}) has also been observed, where irradiated FeS became depleted in S at the surface, forming a Fe-metal cladding; the driver of this mechanism was described as RIS in \cite{Christoph2022} albeit under the synonymous term of `defect driven diffusion', 
 where a maximum diffusion rate is scaled by the number of defects in the sample. 
Similar findings for ion irradiated FeS were identified via X-ray photoelectron spectroscopy \citep{loeffler_laboratory_2008} and TEM/EDS measurements \citep{keller_asteroidal_2013}, while experiments with H- and He-irradiated pentlandite (\ch{Fe_xNi_yS_z}) grains have demonstrated formation of an analogous Fe-rich cladding \citep{chaves_experimental_2025}.
Cation segregation rates of similar values were then assumed to apply to all sulfides based on this troilite (\ch{FeS}) data due to a lack of experimental measurements on non-Fe-containing minerals \citep[SpuBase]{Jaggi2024}. Note, that if segregation of this type happens, a non-stoichiometric sputtering equilibrium is reached, which results in a decrease in S yield by a factor of ten compared to simulations without diffusion, resulting in stoichiometric yields \citep{Jaggi2024}. In this work we will show that neither MgS nor CaS express enhanced sulfur mobilization with irradiation, resulting in S depletion through preferential sputtering, and we discuss the correlation of the ionic spacing in sulfides with their surface depletion behavior to predict the behavior of other sulfides relevant to Mercury.

\section{Methods}

\subsection{Samples}\label{met:samples}
We selected two simple, Mercury-surface-relevant sulfides to elucidate the physics behind the response of Hermean CaS and MgS to ion irradiation. The accumulation of the metal component at the surface of various irradiated sulfides has been previously observed for several compounds including for Ni, Cu, Co, Fe, and Mo sulfides \citep{Coyle1980,Loeffler2008,Feng1974,chaves_experimental_2025}, all except FeS are unlikely to be relevant Mercury analogues \citep[however, Mercury surface Fe:S$\leq$0.09;]{McCoy2018}. 
For this work, an 99.9\% oldhamite (CaS) sample was acquired from Thermo Fisher Scientific (TFS) and two niningerite (MgS) samples from Zegen Metals \& Chemicals Limited (ZMC) and BenchChem (BC).\\
The average bulk composition of the powders was determined by using Scanning Electron Microscopy (SEM) Energy Dispersive Spectroscopy (EDS) on grains deposited on carbon tape. To minimize surface charging and optimize EDS quantification, we worked with a high acceleration voltage (10-15~kV) at low vacuum (0.1~Pa). The sulfur:metal ratio of 0.88 with a standard deviation of $0.01$ was determined for the CaS sample, and 0.75$\pm0.01$ (ZMC) and 0.69$\pm0.03$ (BC) for the two MgS samples respectively. All analyzed powders had 2-3~at.\% of both oxygen and carbon in their bulk analysis. The former is attributed to surface oxidation whereas the latter is mostly attributed to the underlying carbon tape.

\begin{figure}[htb!]
    \centering
    \includegraphics[width=\linewidth]{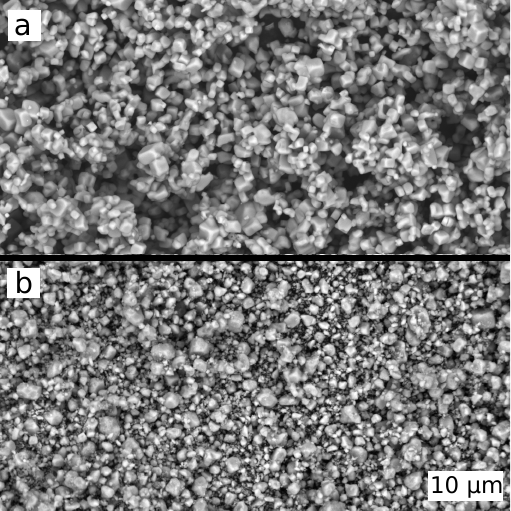}
    \caption{MgS powder a) before and b) after hand milling. \label{fig:Nng_powder_milled}}    
\end{figure}

Initially we worked with MgS from ZMC, however the sample was expected to have suffered extensive atmospheric contamination after being opened during shipping. We used the BC MgS material, which was presumed to be pristine, to explore the effect that the oxidation has on experiment reproducibility and radiation response. Materials from TFS and BC were found to have minimal oxidation indicators; however, we note that these materials do form surface oxides immediately upon atmospheric exposure and must be handled carefully to preserve the surface stoichiometry. In an attempt to reduce the surface oxidation percentage, the original heavily oxidized ZMC MgS powder was milled in an agate mortar before pressing it into the pellet (Fig.~\ref{fig:Nng_powder_milled}). This did not have the desired effect of creating an enhanced amount of fresh surfaces to offset the surface oxidation signal. We attribute this to the small grain size of our initial powders, which when ground do not exhibit a significant increase in the concentration of pristine surfaces due to the high surface to volume ratio.

\subsection{X-ray Powder Diffraction} The overall composition of the sample and the shape of the grains was determined by SEM analysis. To confirm that the crystalline fraction of the sample is $\geq99\%$ sulfide, powder XRD analysis was performed on a Bruker D2 Phaser at University of Virginia. The high crystallinity of the samples resulted in intense diffraction peaks which allowed for a confident matching of the spectral patterns to entries within the Crystallography Open Database \citep{Downs2003,Vaitkus2023}. The Mg and Ca sulfide crystal properties reported in the Discussion (Sec.~\ref{sec:discussion}) were taken from their respective best fit database entry.

\subsection{Pellet creation}
For samples that could be safely mounted in the ion irradiation chamber and easily handled for further investigations, pressed pellets were prepared. The sulfide pellets were created from the powder samples handled only in a nitrogen-purged glove tent and were pressed (for 5 minutes) into 10~mm diameter rings made of 6061-T6~aluminium at 1500~PSI (10.3 MPa). The fresh pellets were then kept under low-vacuum for $\leq120$~minutes before they were mounted onto the analysis platen and transferred into the irradiation chamber under \ch{N2} flow to prevent significant atmospheric contamination.

\subsection{Ion irradiation}
To simulate the solar wind precipitation onto the Hermean-analogue sulfides, we performed an ion irradiation at near normal incidence ($\upalpha_{in} = 15^\circ$) under ultra-high vacuum conditions (10$^{-9}$ Torr) with either \ch{H2+} or \ch{He+} ions at average solar wind speeds (1~keV/amu) utilizing adapted instrumentation at the University of Virginia. The sample surface was neutralized during irradiation using an electron flood gun. Conditions were similar to previous works \citep{Dukes1999,Loeffler2009,laczniak_characterizing_2021,Christoph2022,Chaves2023}. 
The applied ion flux ranged from $2-4\times10^{13}$~ions~cm$^{-2}$~s$^{-1}$ to reach fluences of up to $3\times10^{19}$~ions~cm$^2$, which simulates an exposure time of $\sim$25000~years for Mercury, assuming a mean proton precipitation flux of $3.7\times10^7$~cm$^{-2}$~s$^{-1}$ \citep{Raines2022}. The applied fluxes are in agreement with the fluxes applied in \cite{Christoph2022}, but note that in both cases a large fluence was necessary to remove atmospheric surface contamination and oxidation products not expected for Mercury. The fluence equivalent exposure time is thus not to be taken as an exact measure for the response of sulfides on airless bodies.

\subsection{X-ray photoelectron spectroscopy}
The fluence dependency of the surface composition and the chemical bonds therein were determined by conducting in situ X-ray photoelectron spectroscopy (XPS) spectra at regular, logarithmically spaced fluence steps using a PHI Versaprobe~III XPS microprobe with a monochromatic Al~K$\alpha$~X-ray source (1486.6~eV). 
For the surface composition information, we averaged ten survey spectra that were each taken with a step size of 0.4~eV (except in the He irradiation on CaS, where the step size was set to 1~eV), 
and a pass energy of 224~eV (energy resolution of 2.2~eV). The surface was neutralized during XPS acquisitions using low-energy ions and electrons. For the chemical state analysis we used a step size of 0.1~eV (0.4~eV for He on CaS) and a pass energy of 26~eV (instrument energy resolution: 0.25~eV) instead and averaged over 10--40 scans depending on the intensity of the features. 
The spectral fitting was performed in PHI Multipak v9.8 which includes the instrument-specific sensitivity factors for quantitative analysis. The Shirley model \citep{shirley_effect_1972} was used to remove the inelastic background. Peaks are fitted using Gaussian-Lorentzian curves.
For the MgS (BC), the first measurement has been published as a spectral reference, including detailed acquisition information \citep{jaggi_magnesium_2025}.

\subsubsection{XPS chemical state analysis}
The key advantage of XPS in materials characterization lies in its ability to reveal chemical bonding through shifts in core-level photoelectron peak positions caused by changes in valence electron configuration \citep{sokolowski_chemical_1958,hagstrom_electron_1964,Greczynski2022}. Increased negative charge density raises photoelectron kinetic energy, lowering binding energy. This shift, known as the chemical shift \citep{siegbahn_esca_1967,fahlman_electron_1966}, reflects changes in the chemical environment and informs on the different bonds present at the sample surface.\\
Note that element peak names have a suffix of $nl_j$ where n is the principal quantum number, l is the orbital angular momentum quantum number (orbital l = 0, 1, 2, ... n-1 is denoted with the orbital shell names s, p, d, f). For peaks that originate from outside the s-shell ($l\geq1$) the spin projection ($s\pm1/2$) of the ejected photoelectron is used to determine the total angular momentum quantum number $j = l + s$. It follows that, e.g., the O~1s feature is comprised of one peak per chemical state whereas the Ca 2p and S 2p features have one doublet (2p$_{3/2}$–2p$_{1/2}$) for each chemical state.

\subsubsection{XPS composition analysis}
The sulfide pellet surface composition was determined using the 1s peaks of O, C, and F with the 2p peaks of Ca, Mg, and S. The experimental variation was determined by calculating the average and standard deviation (SD) concentration for each element measured in two consecutive survey spectra at the end of the irradiation experiment. To obtain a sensitivity error, we determined the variation of the surface composition based on compositions resulting from using 2s photoelectron peaks for Ca, Mg and S, instead of the 2p features. The proximity of the 2s peaks of Ca, Mg, and S to their corresponding 2p peaks gives confidence that we sample a comparable information depth within a few nm \citep[e.g.][]{Greczynski2022}. For Mg/Ca and O, we report the SD of the result at the final irradiation step, whereas C and F that are removed with irradiation are evaluated at the first irradiation step. A large error at the final irradiation step therefore reflects the inhomogeneity of the altered sample surface. The reported error is equal to either the experimental error or sensitivity error, whichever is larger.

\subsubsection{XPS energy calibration}\label{met:calibration}
To compare the XPS results with other works, the primary C~1s peak was attributed to C-H bonding in adventitious carbon and set to  284.8~eV \citep[e.g.][]{Gengenbach2021} fixed to the vacuum level. Note that this is not absolute charge referencing due to the issues of using Adventitious Carbon (AdC) as a reference outlined by \cite{Greczynski2020,Greczynski2022}, but it provides a standard referential offset for variation in surface potential between samples. We also observed a fluence-dependent shift in AdC which was comparable to the one observed in \cite{Christoph2022} and is discussed in Appendix~\ref{app:carbon}. Because of the shift and removal of surface AdC, we used the pre-irradiation position of the largest S~2p peak ($\sim160$~eV) as a relative reference to eliminate variations in neutralization efficiency in between fluence steps. We are confident that this calibration allows us to evaluate the change in chemical states of all elements because of the consistent shapes and positions of the observed S~2p sulfide, sulfite, and sulfate peaks.



\subsection{Binary collision approximation simulations}
To reproduce the experimental data and infer relative sputtering yields, we used the Binary Collision Approximation (BCA) Monte Carlo code SDTrimSP to simulate the effects of ion irradiation on sulfides covered by an overlayer of adventitious carbon and oxygen. To obtain good statistics we modeled $\sim4.8\times10^7$ impactors at fluence steps of $6\times10^{14}$~ions~cm$^2$. For binding energies and densities, default settings for SDTrimSP were used. Furthermore, 1~keV \ch{H} was used to simulate the 2~keV \ch{H2+} irradiation under the common assumption that \ch{H2+} will dissociate upon approach to the surface, resulting in two hydrogen atoms with a kinetic energy of 1~keV each.

To obtain a XPS-like signal with similar information depth from SDTrimSP data for direct comparison, a Beer-Lambert-type weighting method was applied to the layers of the simulated target. Photoelectron inelastic mean free path (IMFP) calculations were derived from the NIST electron IMFP database \citep{NIST2010} using the TPP-2M equation and a band gap of 2.7~eV for MgS and 4.434~eV for CaS respectively \citep{stepanyuk_electronic_1989,stepanyuk_electronic_1992}. For CaS, the band gap is given as a range between 2.143--4.434~eV \citep{kaneko_new_1988,stepanyuk_electronic_1992}, however using the lower band gap energy for CaS did not lead to a significant enough difference in the calibration result to warrant a discussion here.\\ 
For plotting purposes, the surface composition data was down-sampled by selecting every tenth data point. To smooth minor statistical variation in the modeled results, a rolling average with a window size of three was applied with a minimum window size of one to preserve the few data points available at fluences $\leq10^{17}$~ions~cm$^{-2}$. 

\section{Results}
\subsection{Surface composition under ion irradiation}
\begin{figure*}
    \centering
    \includegraphics[width=0.48\linewidth]{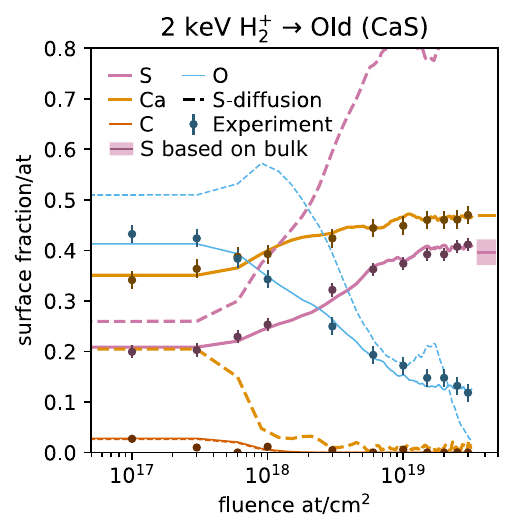}
    \includegraphics[width=0.48\linewidth]{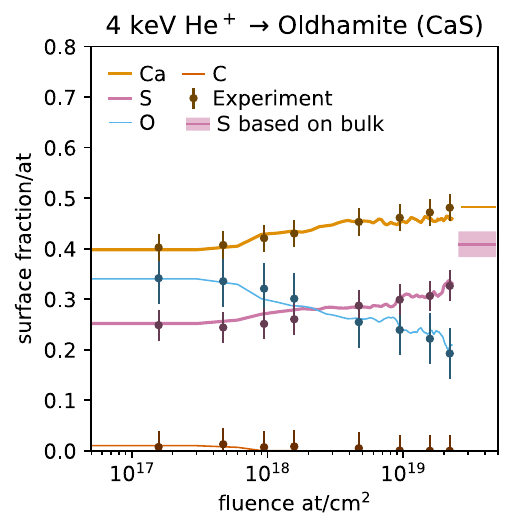}
    \includegraphics[width=0.48\linewidth]{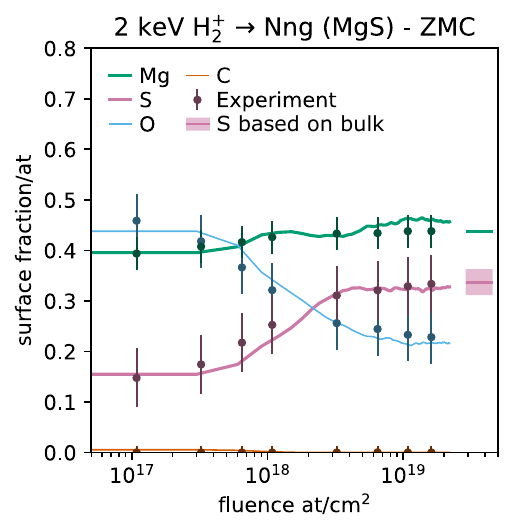}
    \includegraphics[width=0.48\linewidth]{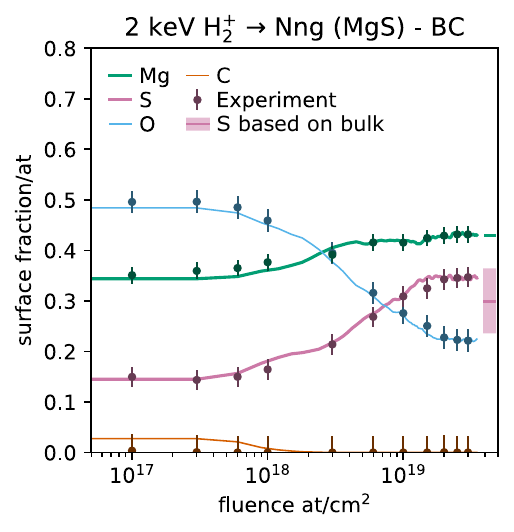}
    \caption{XPS data (Experiment) compared to surface evolution of the corresponding SDTrimSP simulations. The S-diffusion case in the 2~keV \ch{H_2+} on Old (CaS) plot uses the same damage-dependent S diffusion rates determined for the radiation induced segregation in FeS \citep{Christoph2022}. Measurement errors are two standard deviations of the composition obtained by using either 2p or 2s peaks of Ca, Mg, and S for analysis or the experiment error, whichever is larger. For C, the F error of $\pm0.2$~at\% was used (F$\leq0.5$~at\%, not shown). Final step is compared to the sulfur:metal ratio of the bulk (with a 2SD error) using the final measurement's Mg/Ca content.}
    \label{fig:comp_results}
\end{figure*}

After irradiating with $\sim10^{19}$~ions~cm$^2$ onto the CaS and MgS samples the resulting surface S:Mg and S:Ca ratios coincide within error to the bulk sulfur:metal ratios. The exception is the He on CaS case, where after $2.2\times10^{19}$~\ch{He+}~cm$^2$ the surface sulfur content ends up lower than that inferred from the bulk (Fig.~\ref{fig:comp_results}). 
The initial surface greatly differs from the bulk composition due to oxidation and AdC. Initial AdC---typically less than one monolayer---is removed quickly, however the oxidation seems pervasive throughout the sample, likely forming a ``rind" around each individual grain, as well as at the pellet surface. In all irradiation experiments there is a significant amount of surface-bound oxygen ($\geq10$~at\%) which is not removed at the final fluence step. 

\subsubsection{Chemical state analysis}\label{res:chemstate}
On the fresh surface we see characteristic binding energy peaks due to surface oxidation which get removed/reduced during irradiation. For example in the CaS cases, there are three S~2p doublets, one for the S in the native metal sulfide (CaS), and one each for the S bound to sulfates (\ch{CaSO4}) and sulfites (\ch{CaSO3}), respectively, the latter of which get reduced under ion irradiation. Due to the greater electronegativity of atomic oxygen compared to sulfur, the sulfite and sulfate peaks occur at higher binding energies than the sulfide feature. The same is true for the atomic sulfur S~2p feature in MgS, where the electronegativity of an S-S bond (2.5 on pauling scale) lies above that of the Mg-S bond (1.2).

\begin{figure*}
    \centering
    \includegraphics[width=0.8\linewidth]{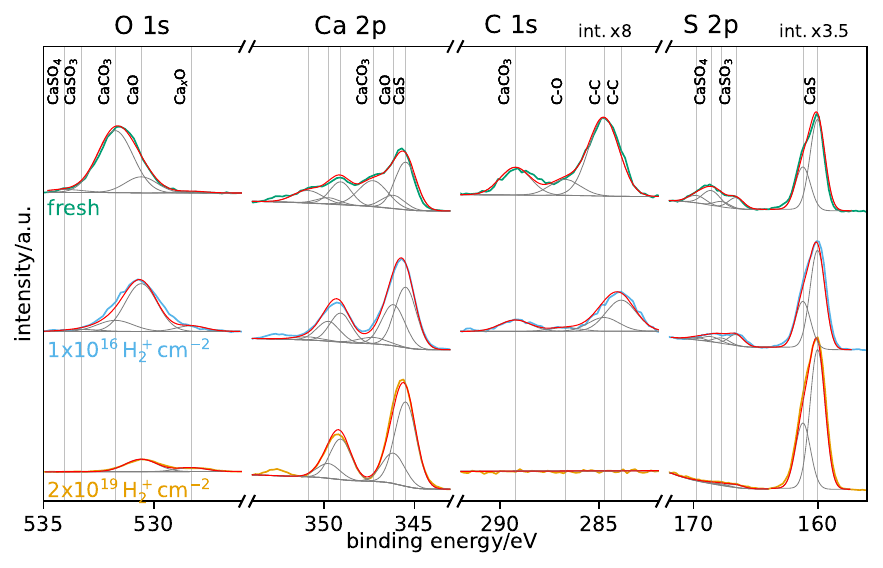}
    \caption{XPS spectra of CaS binding states compared between fresh sample, first He irradiated fluence step ($10^{16}$~\ch{H2+}~cm$^{-2}$) and final fluence step (2$\times10^{19}$~\ch{H2+}~cm$^{-2}$). In the Ca~2p the sulfate and sulfite features are not fitted due to their low abundance, as apparent in the S~2p feature.
    }
    \label{fig:chemical_states_H_CaS}
\end{figure*}

In all experiments we observed that after the removal of C and the majority of O, the S is present primarily bonded with the metal cation; thus, in all high-resolution XPS spectra and at all fluences, the native sulfide constituent chemistry can be discerned. In the \ch{H2+} on CaS experiment (Fig.~\ref{fig:chemical_states_H_CaS}) there is surficial \ch{CaCO3}, formed either during the synthesis process, transport, or handling, that is convolved with the CaS features in the Ca~2p, C~1s, and O~1s peaks before irradiation; the carbonate gets rapidly removed after the first $1e16$~ions~cm$^{-2}$. In the initial Ca~2p signal we only fit the major observable and differentiable peak components of \ch{CaCO3}, \ch{CaO}, and \ch{CaS}. In a similar fashion, sulfates ($\sim$168~eV) and sulfites ($\sim$166~eV) are visible in the S~2p signal along with the native CaS, but are not explicitly fit in the multi-chemistry O~1s peak. Furthermore, no \ch{H2O} or hydroxide was observed, based on measured O~1s photoelectron peak binding energy(s). The contribution of \ch{CaSO4} before irradiation exceeds that of \ch{CaSO3}, but this trend is then inverted, likely due to preferential sputtering of O by the ion irradiation \citep[e.g.,]{Chakrabarti1992}.  

\begin{figure*}
    \centering
    \includegraphics[width=0.8\linewidth]{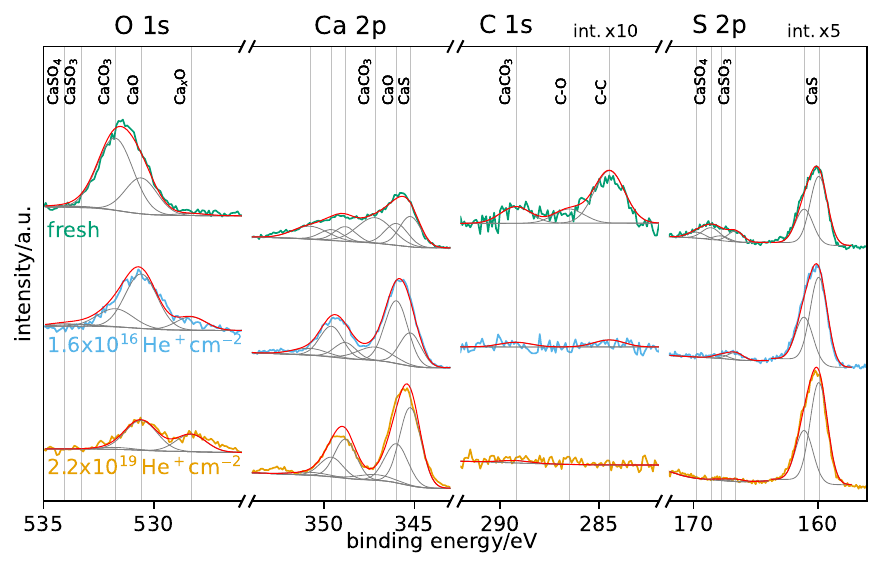}
    \caption{XPS spectra of CaS binding states compared between fresh sample, first He irradiated fluence step (1.6$\times10^{16}$~\ch{H2+}~cm$^{-2}$) and final fluence step (2.2$\times10^{19}$~\ch{H2+}~cm$^{-2}$). In the Ca~2p the sulfate and sulfite features are not fitted due to their low abundance, as apparent in the S~2p feature.}
    \label{fig:chemical_states_He_CaS}
\end{figure*}

 
All features in the \ch{He+} on CaS chemical state data could be fitted using the same binding-energy corrected positions from the proton irradiation case (Fig.~\ref{fig:chemical_states_H_CaS}~\&~\ref{fig:chemical_states_He_CaS}). As expected, there are no differences in the species present at the surface between the \ch{H2+} and \ch{He+} irradiation, as expected. Quantitatively, there is a larger amount of CaO and \ch{Ca_xO} relative to the CaS peak S~2p (note the x5 magnification of the S~2p counts in the He on CaS case). This is linked to the greater oxidation after months of storage following the initial handling of the powder. Note that in neither CaS dataset did we fit the contribution of \ch{CaSO3}, \ch{CaSO4}, and \ch{Ca_xO} (expected around a BE of 350~eV) to the Ca~2p feature due to their small quantities.

\begin{figure*}
    \centering
    \includegraphics[width=0.8\linewidth]{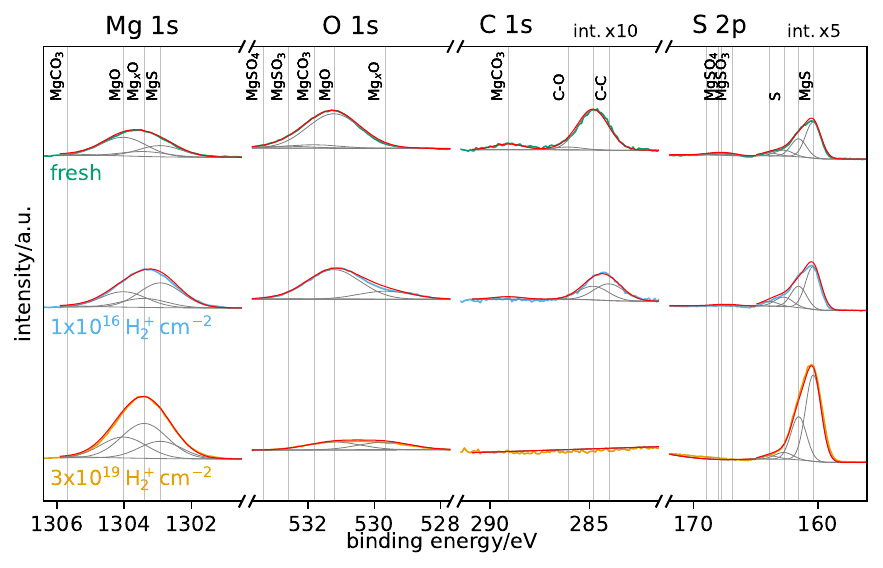}
    \caption{XPS spectra of MgS binding states compared between fresh sample, first proton irradiated fluence step ($10^{16}$~\ch{H2+}~cm$^{-2}$) and final fluence step (3$\times10^{19}$~\ch{H2+}~cm$^{-2}$). In the Mg~1s the sulfate and sulfite features are not fitted due to their low abundance, as apparent in the S~2p feature.}
    \label{fig:chemical_states_MgS}
\end{figure*}

The BC MgS results generally express the same trends as those described for CaS (Fig.~\ref{fig:chemical_states_MgS}). The one difference is the presence of an additional sulfur doublet in the S~2p spectra indicating an incomplete reaction in the synthesis of the MgS. The elemental sulfur is likely low in quantity and/or confined to the grain surfaces because it does not show up in the XRD spectra we performed on the MgS powder.





\section{Discussion} \label{sec:discussion}
We observe that for average solar wind speed protons a sulfur depletion does not occur in neither MgS nor CaS unlike in FeS \citep{loeffler_laboratory_2008,Christoph2022} and \ch{(Fe,Ni)9S8} \citep{chaves_experimental_2025}. This is demonstrated in Figure~\ref{fig:comp_results}, where we applied the same damage-dependent S diffusion rates determined for the RIS in FeS by \cite{Christoph2022} to the 2~keV \ch{H_2+} on CaS simulation.
Although the results of He on CaS and H on MgS are not quite ideal as they show signs of significant surface oxidation, we will now show how the acquired data of oxidized sulfides is still conclusive regarding S mobility in Ca and Mg sulfides before proposing how anion spacing may be responsible for vacancy and S mobilization. 

\subsection{Oxidation layers and relevance of non-stoichiometric data}\label{dis:oxy}
Working with metal sulfides brings great challenges. For MgS it has been shown that all commercial powders exhibit oxidized rims which occur even on freshly produced MgS \citep{kimura_laboratory_2005}. In our laboratory, the pristine Ca and Mg sulfide powders were always kept under a protective atmosphere and only briefly opened in a purged glove tent to extract a few milligrams of sample. We note that the relative humidity in the purged atmosphere could not be reduced below 8\%; the remaining \ch{H2O} can react together with residual atmospheric \ch{CO2} and the sulfides to create \ch{H2S} gas and an oxide layer comprised of Ca/Mg-carbonate, following
\begin{align}
\ch{
MS + CO2 + H2O → MCO3 + H2S
}
\end{align}
Similar processes then form the observed sulfites, sulfates, and oxides. The resulting surface metal to sulfur ratio is thus reflective of the degree of S depletion by these oxidation reactions. This becomes apparent in the 4~keV~He$^+$ on CaS results (Fig.~\ref{fig:comp_results}).

In the irradiation experiments of Co, Ni, Fe and Zn sulfides, \cite{Coyle1980} found an inverse correlation between the susceptibility of the surface to metal reduction and oxidation. 
The rapid and intense surface oxidation of our MgS and CaS samples compared to the resistance towards surface reduction supports this inference.

\subsection{The Effect of Surface roughness on surface oxide removal} 
In the laboratory data, the fluxes of H and He were comparable and the surface compositional evolution of the two CaS samples under irradiation behaved near-identically in time. This is unexpected, because He ions are expected to `clean' the surface of contaminants and oxides at rates of about an order of magnitude greater than \ch{H} ions according to the known difference in sputter yields \citep[e.g., $Y_{He}$:$Y_H\sim$0.07:0.007 for Si;][]{yamamura_energy_1996}. This difference in yield therefore must be balanced by an approximately tenfold increased thickness of the oxidized surface layers for the CaS in the He irradiation case. We find this to be the case using SDTrimSP, where the model data implies a significantly thicker oxide ($\sim700$~nm) for the He irradiated CaS sample compared to the H irradiation ($50-70$~nm). We would expect a thicker layer to have formed between the initial (\ch{H+}) and consecutive (\ch{He+}) CaS experiments because the latter seems more heavily oxidized due to the low final surface sulfur:metal ratio after continuous irradiation, which suggests a higher oxide to sulfide ratio. We would expect the overall thickness of either layers to be thinner than those calculated by SDTrimSP however, because a) realistic binding energies are generally larger than the tabulated ones in this work \citep[e.g.,][]{jaggi_new_2023} and b) surface roughness significantly reduces yields \citep[e.g.,][]{Biber2022}.


Unlike previous work on FeS in \cite{Christoph2022} where thick sections were used, we selected powder targets as a more comparative regolith simulant. However, the intrinsic surface roughness and porosity of the pressed powder pellet complicates the removal of oxides by sputter re-deposition of material from one grain to an adjacent grain; this cannot be reproduced in the 1D version of SDTrimSP used in this work. Therefore, the model layer input information is only representative of the individual initial grain sample composition, but not of the actual sample oxidation depth. Additionally, oxidation of the power grains between experiments, even with significant precautions to minimize atmospheric exposure, further complicates the picture. For example,  
in the case of the H on CaS experiment we expect that the remnant 15~at\% of surface oxygen at a fluence of $\geq$10$^{19}$~ions/cm$^2$ is a result of shadowing in the inter-granular space in the pressed powder sample, which suppresses the removal of oxygen from the sample surface. In the case where the bulk metal to sulfur ratio is not reached with continued irradiation, shadowing is unlikely to solely account for the observed high oxygen content post-irradiation. The shallow sampling depth of XPS ($\leq$30~nm) compared to the grain size (10~$\upmu$m) instead suggests a thicker metal-oxide surface layer which formed from atmospheric exposure (\ch{O2}, \ch{CO2}, and \ch{H2O}), which keeps growing by in-diffusion, and shields the sulfide and sulfur atoms during sputtering (Sec.~\ref{dis:oxy}). However, this oxidation ``rind" does not impact the finding that S in MgS and CaS does not preferentially deplete, nor does it change the mobility of S in CaS and MgS, based on the XPS chemical state data.

\subsection{Chemical states}
We found that the fluence-dependent behavior of the S~2p peak shape can be used as a diagnostic indicator of the behavior of the corresponding Ca or Mg sulfide with irradiation. For CaS, the S~2p feature is made up of a single doublet (2p$_{3/2}$ and 2p$_{1/2}$) assigned to CaS with only a minute  contribution of \ch{CaSO3} in the He irradiation case (2p$_{3/2}$ peak area of \ch{CaSO3} is 2.3\% of S~2p$_{3/2}$ peak) and below detection in the H irradiation case at the final irradiation step (Figs.~\ref{fig:chemical_states_H_CaS}~\&~\ref{fig:chemical_states_He_CaS} ). The MgS experiments are comparable, but with a lesser sulfate component ($<0.2\%$ of S~2p$_{3/2}$ peak) and a minor contribution of elemental sulfur ($7.8\%$ of S~2p$_{3/2}$ peak; Fig.~\ref{fig:chemical_states_MgS}). We attribute the latter (elemental sulfur) to residual reactant from the synthesis procedure. We are thus confident that---after oxide removal---the surfaces exposed to ion irradiation are those of CaS and MgS, respectively. The lack of S depletion thus rejects the possibility of S mobilization in CaS and MgS akin to the the mobilization observed in Co, Ni, and Fe sulfides \citep{Coyle1980,Christoph2022}.\\
We also note that unlike the carbonates, sulfites, and sulfides, which are removed after $10^{16}$~ions/cm$^2$ the oxides are the most resistant to sputter removal. This is due to chemisorption with atoms/molecules present on the surface and supported by the increasing signal of a Ca-undercoordinated oxide with increasing fluence (\ch{Ca_xO} in Figs.~\ref{fig:chemical_states_H_CaS}~\&~\ref{fig:chemical_states_He_CaS}).

\subsection{Anion segregation and metal formation}\label{sec:segregation}

The surface depletion of an element is caused by the preferential transport of a target atom hopping between lattice site vacancies toward the surface, where they are subsequently removed via sputtering or desorption. In this case, vacancy concentrations are enhanced in the near-surface region by increased atomic displacement with sputtering. On the other hand, atoms transported preferentially through interstitial regions tend to trap and accumulate in disordered materials \citep{piller_radiation-induced_1978,lam_bombardment-induced_1986}. For example, \cite{lee_auger_1984} observed for heavily sputtered Fe sulfides that the surface reconstructs by recombination of Fe and sulfur sourced from the bulk if heated to $450^\circ$C for 5 minutes (annealing), enhancing the mobility of S in FeS. Based on this observation we expect sulfur to be transported to the surface via vacancies, where these atoms will be removed via sputtering. We will now highlight some criteria that were proposed to predict segregation before introducing our own criterion, based on the results of this and prior work in Section~\ref{dis:criterion}.

\subsubsection{Enthalpy of atomization}\label{met:atomize} 

In an attempt to predict the segregation behavior of S in MgS and CaS, we refer to a chemical bonding hypothesis proposed in \cite{parker_ion-impact_1975} \citep[and reviewed in ][]{Naguib1975}. 
The authors suggested that the dissociation reaction associated with the lowest heat of atomization $H_a$ per gas atom in a reaction, and therefore the lowest energy barrier, would reflect the behavior of the surface under irradiation. In the model, the compound is broken into its atomic constituents by the energy transfer deposited along the ion implantation track (thermal spike) during irradiation, which may atomize the most volatile constituent(s) (e.g., S) to their gas forms, in addition to bond-breaking and atomic displacement in the solid state. For the example of FeS, that would be:
\begin{equation}
\label{eq:atomize}
    \begin{split}
    \ch{FeS(s)} &\rightarrow \ch{Fe(s)} + \ch{S(g)},\: H_a \approx 3.64~\text{eV/gas atom} \\
    &\rightarrow \ch{Fe(g)} + \ch{S(g)},\: H_a \approx 3.97~\text{eV/gas atom}.
    \end{split}
\end{equation}
The minima of the energies for the end products in this ``enthalpy of atomization" hypothesis is predicted to be indicative of the final state of the solid, assuming ion fluences beyond equilibrium, resulting in either a stoichiometric surface or one depleted in the most volatile constituent. This approach was widely successful when applied to 38 of the 41 cases of elemental enrichment effects with irradiation on diatomic targets reported in \cite{Naguib1975}, with a 100\% accuracy for sulfides. It was, however, already noted in the  \citeauthor{Naguib1975} review that this agreement may be coincidental but effective.

We find that the $H_a$ per gas atom for MgS and CaS at 300~K, the lattice temperature prior to the energy spike \citep{parker_ion-impact_1975}, calculated following Eq.~\ref{met:atomize}, are 7.99 and 9.62~eV/gas atom for complete atomization of the solid sulfide respectively. The comparative value for atomization of the S with the preservation of a metallic Mg/Ca solid surface layer is 3.23 and 3.89~eV/gas atom, respectively. The enthalpy of atomization theory, which suggests the lowest energy reaction is preferred, would therefore erroneously predict that both of our irradiated sulfides would form a metallic phase, and possibly a metal cladding.

\subsubsection{Lattice constant of metal and substrate}
Another explanation for why certain---but not all---diatomic molecules may develop a cation-enriched surface layer was proposed by \cite{bennewitz_bulk_1999} based on Ca metal colloid formation after electron sputtering of \ch{CaF2} \citep{reichling_surface_1996}. 
In this material, the F was observed to diffuse outwards of the irradiated \ch{CaF2} crystal, creating some subsurface F-filled vesicles. The colloids then form to stabilize the surface by aggregation of F-centers (anionic vacancy filled with unpaired electron(s)) that were created within the bulk. To explain why \ch{CaF2} forms Ca metal colloids, but \ch{CaO} does not, \citeauthor{reichling_surface_1996} proposed a ``favorable match" between lattice constant and volume of the Ca metal and Ca fluoride. This criteria falls flat for CaS because the atomic spacing of CaS used in this work (5.70~\AA{}) is comparable to the spacing of \ch{CaF2} (5.46~\AA{}) and Ca metal (5.57~\AA{}) given in \cite{bennewitz_bulk_1999}, unlike the radiation-hard CaO with a lattice constant of 3.83~\AA{}. For this reason we propose a different bond-property that might dictate on the behavior of sulfides under ion irradiation.

\subsubsection{Anionic spacing as a proxy for radiation hardness}\label{dis:criterion}

As a proxy for the mobility of vacancies we have compiled the shortest anionic (S-S) spacing of various sulfides with known and unknown irradiation response (Tab.~\ref{tab:sulfides_lit}). This data was sourced from observations in irradiation literature \citep{Coyle1980,Tsang1979,Loeffler2008,Naguib1975,klein_understanding_2017,Baker1999} and experimentally observed structure files available on the Materials Project website at materialsproject.org. All sulfides that were not observed to show RIS with consequent cation surface enrichment express an S-S anionic spacing exceeding $\sim$3.2~\AA{} within their ideal, crystalline structures, which we assume to limit S-vacancy mobility. 
Returning to the example of Ca colloid formation, \ch{CaF2} expresses an F-F spacing of 2.76~\AA{}, a distance below our anionic spacing cutoff, which would accurately predict metal formation. CaO has an anionic spacing of 3.4~\AA{} and CaS one of 4.03~\AA{} (and MgS one of 3.68~\AA{}), which would accurately predict the absence of colloid formation by hindering anionic vacancy diffusion. 
This predictive model is meant to motivate further research into the stability of sulfides. The predictive model reaches its limit if anion distances within a mineral lie at the presumed cutoff distance. For ZnO for example, an oxide known to be radiation hard \citep[e.g.]{azarov_extended_2017}, we found that anion distances range between 3.06-3.22~\AA{}, depending on the crystal structure. There are however other factors that can affect radiation-hardness, which we will discuss in the context of Mercury's surface.

\begin{table*}
\centering
\caption{Sulfide experimental results and bond lengths.\label{tab:sulfides_lit}}
\renewcommand{\arraystretch}{0.8}
\resizebox{\textwidth}{!}{%
    \begin{tabular}{l ll ll cc ll}
          & beam & method & crystal system &Space Group&min(S-S)/\AA&min(M-S)/\AA&id$^a$& ref. \\
        \hline
        \multicolumn{9}{c}{surface metal formation}\\
        \hline
        NiS$^\dagger$& 1.2 keV Ar+ & XPS &-&-&-&-& - & \citep{Coyle1980}\\
        &&&Hexagonal&P6$_3$/mmc&3.26&2.37&mp-594& \\
        &&&Trigonal&R3m&3.32&2.25&mp-1547& \\
        CuS$^{\dagger,h}$ & 1.2 keV Ar+ & XPS &-&-&-&-& - & \citep{Coyle1980}\\
        &&&Orthorhombic&Cmcm&2.09&2.09&mp-555599& \\
        &&&Hexagonal &P6$_3$/mmc&2.09&2.09&mp-504& \\
        &&&Hexagonal&P6$_3$/mmc&3.09&2.47&mp-558139& \\
        CoS$^\dagger$  & 1.2 keV Ar+ & XPS &-&-&-&-& - & \citep{Coyle1980}\\
        &&&Hexagonal&P6$_3$/mmc&3.21&2.32&mp-1274& \\
        FeS$^\dagger$  & 1.5 keV Ar+ & XPS &-&-&-&-& - &\citep{Tsang1979}\\
        & 1.5 keV Ar+ & XPS &-&-&-&-& - & \citep{Loeffler2008}$^f$\\
        &&&Hexagonal&P6$_3$/mmc&3.47&2.43&mp-2099& \\
        &&&Hexagonal&P$\overline{6}$2c&3.23&2.28&mp-2779& \\
        &&&Monoclinic&P12$_1$/c1&2.88&2.14&mp-22652& \\
        &&&Orthorhombic&Pnma&3.26&2.22&mp-21410& \\
        &&&Tetragonal&P4/nmm1&3.52&2.17&mp-505531& \\
        &&&Hexagonal&P6$_3$mc&3.07&2.26&mp-616476& \\
        \ch{FeS2}$^\dagger$ & 1.5 keV Ar+ & XPS &-&-&-&-& - & \citep{Tsang1979}\\
        &&&Cubic&Pa$\overline{3}$&3.07&2.26&mp-226& \\
        &&&Orthorhombic&Pnnm&3.11&2.23&mp-1522& \\
        \ch{MoS2}$^\dagger$ & 300 eV Ar+ & AFS &-&-&-&-& - & \citep{Feng1974}$^e$ \\
        \ch{MoS2}$^\dagger$ & - & - &-&-&-&-& - & \citep{Naguib1975} \\
        &&&Trigonal&R3m&3.10&2.41&mp-1434&\\
        &&&Hexagonal&P6$_3$/mmc&3.19&2.42&mp-2815& \\
        &&&Hexagonal&P6$_3$/mmc&3.10&2.41&mp-1018809& \\
        &&&Trigonal&R$\overline{3}$m&3.19&2.44&mp-558544& \\

        \multicolumn{9}{c}{}\\
        \multicolumn{9}{c}{\textbf{Prediction}}\\
        \hline
        MnS$^\ddagger$ & &  &Cubic&Fm$\overline{3}$m&2.58&3.65&mp-2065& \\
        &&&Cubic&F$\overline{4}$3m&2.14&3.50&mp-1783& \\
        &&&Cubic&Pm$\overline{3}$m&3.10&2.68&mp-556853& \\
        &&&Hexagonal&P6$_3$mc&3.95&2.42&mp-2562& \\
        \ch{Mg(FeS2)2}&&&Cubic&Fd$\overline{3}$m1&3.01&2.31&mp-1389421$^d$&\\
        \ch{Mg(FeS2)2}&&&Orthorhombic&Imma&3.22&2.32&mp-1045442$^d$&\\
        \ch{Mg(Fe2S3)2}&&&Trigonal&R$\overline{3}$&3.12&2.25&mp-2217928$^d$&\\
        \ch{Mg(FeS2)4}&&&Trigonal&R$\overline{3}$m&3.17&2.28&mp-1394404$^d$&\\
        \ch{Mg(FeS2)4}&&&Trigonal&R$\overline{3}$m&3.12&2.28&mp-1444514$^d$&\\
        (Fe,Mg)S &&& -&-&-&-& - & \citep{Renggli2023}\\
        \ch{Fe(NiS2)2} & 1 keV H+ & XPS &-&-&-&-& - & \citep{chaves_experimental_2025}\\
        \ch{Fe(NiS2)2} & 1 keV H+ & XPS &Cubic&Fd$\overline{3}$m&3.18&2.12&mp-505522& \\
        \hline
        \multicolumn{9}{c}{no detected surface metal}\\
        \hline
        \ch{Li2S}$^\dagger$ & - keV Ar+ & XPS &-&-&-&-& - & \citep{klein_understanding_2017}$^g$ \\
        &&&Cubic&Fm$\overline{3}$m&4.00&2.46&mp-1153& \\
        &&&Orthorhombic&Pnma&3.79&2.39&mp-1125& \\
        \ch{MoS2}$^\ddagger$ & 3 keV Ar+ & XPS &-&-&-&-& - & \citep{Baker1999}$^e$ \\
        &&&Hexagonal&P6$_3$/mmc&3.19&2.42&mp-2815& \\
        ZnS$^\dagger$  & 1.2 keV Ar+ & XPS &-&-&-&-& - & \citep{Coyle1980}\\
        &&&Cubic&F$\overline{4}$3m&3.81&2.33&mp-10695&\\
        &&&Trigonal&P3m1&3.81&2.33&mp-554820& \\
        MgS$^\ddagger$ & 2 keV/amu \ch{H2+} & XPS & cubic &Fm$\overline{3}$m&3.68&2.60&mp-1315$^b$& this work\\
        MgS$^\ddagger$ & 4 keV \ch{He+} & XPS & cubic &Fm$\overline{3}$m&3.68&2.60&mp-1315$^b$& this work\\
        CaS$^\ddagger$ & 2 keV/amu \ch{H2+} & XPS & cubic &Fm$\overline{3}$m&4.03&2.85&mp-1672$^c$& this work\\
        CaS$^\ddagger$ & 4 keV \ch{He+} & XPS & cubic &Fm$\overline{3}$m&4.03&2.85&mp-1672$^c$& this work\\
        \multicolumn{9}{c}{}\\
        \multicolumn{9}{c}{\textbf{Prediction}}\\
        \hline
        TiS$^\dagger$ &&  &Hexagonal&P6$_3$/mmc&3.27&2.48&mp-554462& \\
        &&&Trigonal&R$\overline{3}$m&3.45&2.37&mp-557762& \\
        &&&Hexagonal&P$\overline{6}$m2&3.27&2.48&mp-1018028& \\
        \ch{TiS2}$^\dagger$ &&  &Hexagonal&P$\overline{3}$m1&3.42&2.42&mp-2156& \\
        &&  &Cubic&F$\overline{4}$3m&3.40&2.43&mp-9027& \\
        &&  &Monoclinic&C2/m&3.42&2.42&mp-1077263& \\
        &&  &Monoclinic&C2/m&3.42&2.42&mp-1062030& \\
        &&  &Trigonal&R$\overline{3}$m&3.42&2.42&mp-558110& \\
        \ch{CrS}$^\dagger$ &  &  & Hexagonal &P6$_3$mmc&3.39&2.37&mp-523& \\
        \ch{CrS2} &  &  & Monoclinic &C2/m&3.35&2.35&mp-28910& \\
        \ch{Cr2S3} &  &  & Hexagonal &P$\overline{3}$1c&3.30&2.43&mp-13685& \\
        \ch{Cr2S3} &  &  & Rhombohedral &R$\overline{3}$&3.31&2.39&mp-555569& \\
        \ch{Cr3S4} &  &  & Monoclinic &C2/m&3.58&2.37&mp-964& \\
        \ch{CaMgS2} &  &   &Trigonal&R$\overline{3}$m&3.65&2.66&mp-1227049$^d$& \\
        (Ca,Mg)S &  &   &-&-&-&-& - & \citep{Renggli2023}\\
        \hline
        \multicolumn{9}{c}{Unknown behavior}\\
        \hline
        (Fe,Ca,Mg)S &&& -&-&-&-& - & \citep{Renggli2023}\\
        (Ti,Fe,Ca,Mg)S &&& -&-&-&-& - & \citep{Renggli2023}\\
    \hline
    \multicolumn{9}{l}{Table comments:}\\
    \multicolumn{9}{l}{$^\dagger$ Follow enthalpy of atomization trends \citep{Naguib1975}.}\\
    \multicolumn{9}{l}{$^\ddagger$ Do not follow enthalpy of atomization trends \citep{Naguib1975}.}\\
    \multicolumn{9}{l}{$^a$ file id of experimentally observed species sourced from materialsproject.org (mp).}  \\
    \multicolumn{9}{l}{$^b$ coincides with Crystallography Open Database entry 8104342.}  \\
    \multicolumn{9}{l}{$^c$ coincides with Crystallography Open Database entry 9008606.} \\
    \multicolumn{9}{l}{$^d$ Not experimentally observed structure.} \\
    \multicolumn{9}{l}{$^e$ We assume that the higher sputter yields of 3~keV compared to 300~eV \ch{Ar+} prevents accumulation of Mo metal.}\\
    \multicolumn{9}{l}{$^f$ reduction to metallic Fe at fluences above $10^{19}$~ions cm$^{-2}$.} \\
    \multicolumn{9}{l}{$^g$ surface `cleaning' with unknown total fluence - unknown behavior at elevated fluences.} \\
    \multicolumn{9}{l}{$^h$ Highly sensitive to thermodynamic data used. $\Delta H_a$ between full and partial atomization (metallic layer formation) is only 0.04 eV/gas atom.}
    \end{tabular}
    }
\end{table*} 

\subsection{Sulfide behavior on Mercury}

There are major uncertainties regarding the behavior of sulfides when non-endmember sulfide compositions are considered. It is known for ZnO that the presence of implanted contaminants (e.g., B, Ar, and Ag) can act as defect stabilizers, preventing defect annihilation and promoting ``radiation hardness"  \citep{kucheyev_ion-beam-produced_2003,azarov_crucial_2014}. Small amounts of metals, for example: a few at\% of Mg added to CaS have been observed to significantly reduce the sulfide's tendency to oxidize (Christian J. Renggli, pers. comms.). Based on the observed inverse proportionality of oxidation readiness and radiation-hardness, we would thus expect that these lattice substitutions can greatly change the radiation behavior of, e.g., \ch{Mg_xCa_{x-1}S} \citep{Coyle1980}. Note that the complex, Fe-bearing mixtures in Table~\ref{tab:sulfides_lit} all have an anionic spacing situated at the proposed anionic spacing cutoff, which we would assume to express low radiation hardness and therefore surface metal formation. Based on this observation, the behavior of Mercury's sulfides under solar-wind irradiation may be strongly tied to their respective composition. 

One other effect that greatly impacts the irradiation response of sulfides on Mercury is heat. Our experiments were performed at room temperature, however Mercury boasts surface temperatures ranging from freezing 93.15~K to blistering 700~K, which will affect the radiation hardness of materials. With increased temperature, the defect mobility increases and, if heated above the crystallization temperature, the irradiated damaged mineral could be annealed, reforming its original crystal structure by defect recombination or by defects diffusing to the sample surface or other sinks such as gas-filled vesicles or defect tracks \citep{lam_bombardment-induced_1986,silva_nitrogen_2010,azarov_crucial_2014}. 
On the other hand,cryogenic temperatures within the shadowed and polar regions of Mercury, as well as on the Hermean nightside, could cause defects to accumulate to a degree that even pure Mg and Ca sulfides express S segregation if defect production and mobility exceed defect annihilation rates. The precipitation of ions onto Mercury's surface is mostly confined to the magnetospheric cusps on Mercury's dayside \citep[e.g.,]{Raines2022}, therefore we do not expect cryogenic conditions to play a major role in the sulfide sputter interaction. Instead, the high temperatures of Mercury would enhance radiation-hardness of all sulfides on Mercury, possibly even for Fe-bearing sulfides. As a result we would expect near stoichiometric removal of S and metal from Mercury's sulfides by means of solar wind ion sputtering, which should be reflected in the Hermean exosphere composition.

\section{Conclusion}
The surface depletion of S and the formation of a metal cladding in sulfide minerals does not occur in either CaS or MgS under ion irradiation, unlike in Ni, Cu, Co, Fe, and Mo-sulfides. We predict that sulfide species, such as Ti, Cr, and Ca-Mg sulfides, with an anionic S-S spacing $\geq$3.2~\AA{}, will sputter stoichiometrically from the Hermean surface when irradiated by keV solar-wind ions.  Assuming Mercury’s sulfides predominantly behave like our analogue minerals, then we can expect no attenuated S sputter yield nor radiation-induced segregation of S. This implies that the $\sim3.5$~wt\% sulfur present within Mercury's regolith should be directly reflected in similar concentrations within the planetary exosphere. The low glow-factor of sulfur is thus likely the sole reason for MESSENGER's non-detection of S by means of UV-spectroscopy, and the Mass Spectrum Analyzer aboard BepiColombo should provide sufficient mass resolution to detect and distinguish singly-charged sulfur from other atomic ions.

\section*{Open Research Section}

The raw XPS spectra and the compositional data are available on https://doi.org/10.5281/zenodo.15801849; a link to this repository appears on the Laboratory for Astrophysics and Surface Physics  website (https://engineering.virginia.edu/LASP). The pre-irradiation spectra for the BenchChem MgS was published as a spectral reference and includes detailed acquisition information \citep{jaggi_magnesium_2025}.

\begin{acknowledgements}
Financial support has been provided by the Swiss National Science Foundation Fund (P500PT\_217998)), the National Science Foundation Astronomy Program (2009365) and the NASA Solar System Workings Program (12000470-028) for financial support. The authors also acknowledge \hyperlink{https://rc.virginia.edu}{Research Computing at The University of Virginia} for providing computational resources that have contributed to the results reported within this publication. XPS/SEM/EDS/XRF data were acquired at the University of Virginia Nanoscale Materials Characterization Facility (NMCF), supported in part by the School of Engineering and Applied Science. Special thanks go to Diane A. Dickie for the insights on the XRD spectra, Cassandre M. Morel for providing a precursor code for the SDTrimSP-to-XPS signal calibration, and Andreas Mutzke for providing clarifications on the damage-driven diffusion implementation in SDTrimSP. 
\end{acknowledgements}

\begin{appendix}
\section{Behavior of carbon}\label{app:carbon}
We observe a shift to lower BEs of the major C~1s peak by 0.7--0.8~eV between the fresh sample and irradiated sample. This coincides well with the $\sim$0.8~eV C~1s peak shift on FeS shown in the supplementary material Figure~S5 in \cite{Christoph2022}. The extent of the shift coincides with the shift from disordered state carbon  \citep[284.8~eV in ]{Reinke1996,Luthin2000} to more graphitic or sp-hybridized carbon \citep[284.0~eV in ][]{Reinke1996,Luthin2000}. Therefore the C~1s peak of the fresh sample would represent the initially disordered AdC adsorbed on the surface, whereas the graphite-like state would become predominant after irradiation. Other possible explanations for the shifts could be related to coverage and the presence of carbides, but the extent of the observed shift does not agree with those observed in literature. For example, the binding energy dependency on the level of surface coverage of $\leq$0.18~eV reported in \cite{Luthin2000} is not large enough to account for the observations whereas the presence of a carbide would be responsible for a significantly larger shift of $\sim$1.9~eV \citep{Pillai2015}.
\end{appendix}


\clearpage
\newpage
\bibliography{zbibliography.bib}{}
\bibliographystyle{aasjournal}

\end{document}